\documentclass[twocolumn,showpacs,preprintnumbers,amsmath,amssymb,superscriptaddress]{emulateapj}

\newcommand{\Msun}{~M_\odot}

\newcommand{\gcm}{\rm ~g~cm^{-3}}
\newcommand{\kms}{\rm ~km~s^{-1}}
\newcommand{\ergs}{\rm ~erg~s^{-1}}

\newcommand{\ml}{~\Msun ~\rm yr^{-1}}

\usepackage{hyperref}

\begin{document}

\title{EVOLUTION OF THE CRAB NEBULA IN A LOW ENERGY SUPERNOVA}

\author{Haifeng Yang and Roger A. Chevalier}
\affil{Department of Astronomy, University of Virginia, P.O. Box 400325, \\
Charlottesville, VA 22904; hy4px@virginia.edu, rac5x@virginia.edu}


\begin{abstract}

The nature of the supernova leading to the Crab Nebula has long been controversial because of
the low energy that is present in the observed nebula.   
One possibility is that there is significant energy in
 extended fast material around the Crab  but searches for such material have not led to detections.
An electron capture supernova model can plausibly account for the
low energy and the observed abundances in the Crab.
Here, we examine the evolution of the Crab pulsar wind nebula inside a freely expanding supernova
and find that the observed properties are most consistent with a low energy event.
Both the velocity and radius of the shell material, and the amount of gas swept up by the pulsar wind point
to a low explosion energy ($\sim 10^{50}$ ergs).
We do not favor a model in which circumstellar interaction powers the supernova luminosity
near maximum light because the required mass would limit the freely expanding ejecta.

\end{abstract}


\section{INTRODUCTION}

The identification of the supernova type of SN 1054, the event leading to the Crab Nebula, has been an
enduring mystery.
The amount of mass in the Crab filaments, $ 4.6\pm 1.8\Msun$ \citep{fesen97}, and their typical velocity, $1260\kms$ \citep{temim06}, imply
a kinetic energy of $\sim 7\times 10^{49}$ ergs, considerably less than the
$\sim 1\times 10^{51}$ ergs  inferred for typical core collapse supernovae.
One possible reason for the discrepancy is that what we view as the Crab Nebula is only the inner
region of a more extended supernova event, which has a normal energy, $\sim 10^{51}$ ergs \citep{chevalier77}.
A constant density freely expanding region with this energy and mass of $6\Msun$ would extend to $\sim 5300\kms$.
However, there is no sign of such a fast ejecta envelope interacting with the surrounding medium
at X-ray \citep{seward06} or radio wavelengths \citep{frail95}, which suggests that the density of the surrounding medium is low and the fast
ejecta would be in free expansion.
In that case, the freely expanding ejecta are heated and
 photoionized by the radiation from the Crab Nebula \citep{lundqvist86,wang13},
but they have not been clearly detected in either emission lines
\citep{fesen97} or in absorption to high velocities along the line of sight to the central pulsar \citep{sollerman00,lundqvist12,lundqvist14}.

If the surrounding $\sim 10^{51}$ erg supernova remnant is absent, SN 1054 must have been a low energy 
and relatively low mass event, and there is theoretical support for such a view.
\cite{nomoto82} found that the abundances in the Crab Nebula pointed to a
progenitor mass in the range $8-10\Msun$ and a probable electron capture supernova.
Models of electron capture supernovae have shown that neutrino driven explosions
occur, but with a relatively small energy
$\ga 1\times10^{50}$ ergs  \citep{kitaura06}.
On the observational side, low energy (and low velocity) supernovae have been discovered,
including IIP events such as SN 1997D, with an estimated energy of
$10^{50}$ ergs \citep{chugai00}.
However, SN 1997D had a peak absolute magnitude of $-14$, considerably fainter than
SN 1054 at maximum.
\cite{chugai00} suggested that SN 1054 was a low energy supernova, with the luminosity
provided by circumstellar interaction.
\cite{smith13} noted that low velocity, Type IIn events like SN 1994W and SN 2011ht are
likely examples of such events and are potential analogs to SN 1054.
However, \cite{tominaga13} followed the presupernova evolution as well as the explosion of an
electron capture supernova, finding that a peak luminosity comparable to SN 1054 could be produced without strong circumstellar interaction; the high peak luminosity is related to the large extent of the progenitor star.

In the circumstellar interaction scenario for the peak emission from SN 1054, the dense 
filaments in the Crab contain circumstellar material and the medium around the observed
nebulae is the initial surroundings, not supernova ejecta.
An alternate view, which we investigate here, is that the nebula is surrounded by freely
expanding ejecta, as in the model of \cite{chevalier77}, but we allow for a low energy
supernova.
Basic constraints on the model are discussed in Section 2.
Pulsar wind bubble models are presented in Section 3 and the conclusions are in Section 4.
The observations of the Crab are spread over the past decades.  Here we take 1994 as
a typical time of observation so that the age of the Crab Nebula is 940 yr. 
The distance to the Crab is taken to be 2 kpc \citep{trimble73}.

\section{OBSERVATIONAL CONSTRAINTS}

\begin{figure}
\centering
\includegraphics[width=0.5\textwidth]{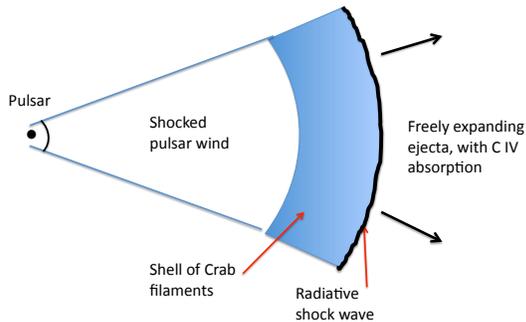}
\caption{Schematic diagram of the model for the Crab Nebula discussed in the text.}
\label{pic}
\end{figure}

The view taken here is that the outer boundary of the observed Crab is the interface of
the pulsar wind nebula with freely expanding ejecta to the outside, as in the model
of \cite{chevalier77}.
The fact that the synchrotron emitting nebula fills the region of
thermal emission is highly suggestive of the outer boundary being the boundary
with freely expanding ejecta.
Additionally, \cite{sankrit97} found that the emission of \ion{C}{4} and \ion{Ne}{5} at the outer edge of the Crab
is consistent with shock emission where the pulsar bubble is driving into freely
expanding ejecta, but is inconsistent with a photoionization model.
The Ne V emission, in particular, appears to come from a ``skin'' to the Crab
\citep{hester08}.
The basic model presented by \cite{sankrit97} has a shock velocity of $v_{sh}=150\kms$
and a preshock density of $\rho_0=12.7 m_H$ cm$^{-3}$, where $m_H$ is the proton mass.  The model requires that
there be a cooling shock wave to produce the line emission, although there may not
be such a shock front around the whole nebula \citep{loll13}.
The situation is illustrated in Figure \ref{pic}.

The outer layers of the main body of the Crab Nebula are expanding at $1700-1800\kms$.
If this is cooled postshock gas, the free expansion velocity at the outer edge
is $1550-1650\kms$ if there is a $150\kms$ shock front.
There is evidence for freely expanding gas extending to $\sim 2500\kms$, well beyond 
the velocity of the Crab shell material.
One piece is the spectrum of the central pulsar, which shows \ion{C}{4} absorption extending
to $\sim 2500\kms$ \citep{sollerman00,lundqvist14}.
The absorption does not show the broad flat trough that would be expected if the Crab were
surrounded by the extended envelope of a normal $10^{51}$ erg supernova.
A pulsar wind nebula is expected to have higher velocities in the shocked wind close to the
pulsar, but there has been no detection of high velocity thermal gas in this region; presumably the
gas has been swept out by the flow of relativistic fluid.

The other piece of evidence comes from the ``jet'' on the north side of the Crab.
\cite{rudie08} found that the emission along the sides of the jet is consistent with freely expanding
gas that has been moving with approximately constant velocity since 1054.  The outer tip of the jet
extends to $\sim 2500\kms$  \citep{rudie08,black15}.
A plausible origin for the jet is that relativistic fluid from the main body of the Crab has pushed out
and moved aside freely expanding gas outside the Crab.
In this way, outward acceleration of the gas could be relatively small.
\cite{rudie08} refer to this picture as the aneurism model.
\cite{sankrit97}  advocate such a model to explain the [\ion{O}{3}] skin at the sides
of the jet.
These considerations suggest a model in which the density is declining outside the main body of the Crab and freely expanding gas extends to  $\ga 2500\kms$.

Another constraint comes from the acceleration of the optical filaments determined by
proper motions.
The angular distance from the center of expansion divided by the proper motion gives the
year of explosion on the assumption of constant velocity:
$1140\pm 15$ \citep{trimble68}, $1120\pm 7$ \citep{wykoff77}, and $1130\pm 16$ \citep{nugent98},
corresponding to $\Delta T$ of $86\pm 15$, $66\pm 7$, and $76\pm 16$ yr, respectively,
where $\Delta T$ is the difference between the actual explosion date (1054) and the constant velocity
explosion date.
These results are for optical filaments that are spread around the extent of the Crab Nebula.
A complication with the Crab is that the main shell extends over a range of velocities.
The outer edge of the nebula is moving at $\sim 1700\kms$ \citep{clark83,sollerman00}, but
the brightest emission shows shell expansion at $1260\kms$ \citep{temim06} and many of
the observed knots are probably in this structure. 
The free expansion velocity at the outer edge is $\sim 1600\kms$.
\cite{bieten91} have  examined the expansion of the outer radio synchrotron emission,
obtaining an explosion date of $1245\pm 92$, or $\Delta T=191\pm 92$ yr.
In the picture of filament formation by Rayleigh-Taylor instability, there is the expectation of higher
acceleration in the outer parts of the nebula, and there is some evidence for such
higher acceleration.

For the model in which the pulsar nebula is expanding into freely expanding ejecta, the
shock velocity can be expressed as $v_{sh}=V_s-(R/t_a)$, where $V_s= dR/dt$ is the velocity
of matter at the outer edge of the nebula in the rest frame and $t_a$ is the true age of the Crab.
For material at the outer edge,
\begin{equation}
\Delta T=t_a-\frac{R}{V_s}=\frac{t_a}{1+(R/t_a v_{sh})}.
\end{equation}
For $v_{sh}=150 \kms$, $R/t_a=1600\kms$, and $t_a=940$ yr, we have $\Delta T=81$ yr.
The acceleration needed to produce the shock velocity in the freely expanding ejecta is
roughly consistent with the observed acceleration.

The SN 1054 progenitor is estimated to have an initial mass of $8-10\Msun$,
based on the observed abundances  \citep{nomoto82}.
Allowing for $1.4\Msun$ going into the central neutron star and $1\Msun$, or more, going
into presupernova mass loss, we have an ejecta mass, $M_{SN}$, of $6.6\pm 1\Msun$, or less if there
is more mass loss.
The current Crab Nebula has a mass $M_{sw} \sim 4.6\pm 1.8\Msun$ \citep{fesen97}.
The ratio of swept up mass to total supernova mass is then $M_{sw}/M_{SN}\approx 0.70\pm 0.30$.
If the observed nebular mass is composed of supernova ejecta swept up by the pulsar wind nebula, as
in the model advocated here, much of the ejecta are likely to have been swept up.

\section{EVOLUTION OF THE CRAB NEBULA}

There are several components of the evolutionary model.
In the presupernova stage, there is mass loss that provides the outer target for the
expanding supernova.
Within days after the supernova, the supernova ejecta approach free expansion.
The outer parts interact with the presupernova mass loss, while the inner parts interact
with an expanding pulsar wind bubble.

\subsection{Approximate Model}

\begin{figure}
\plotone{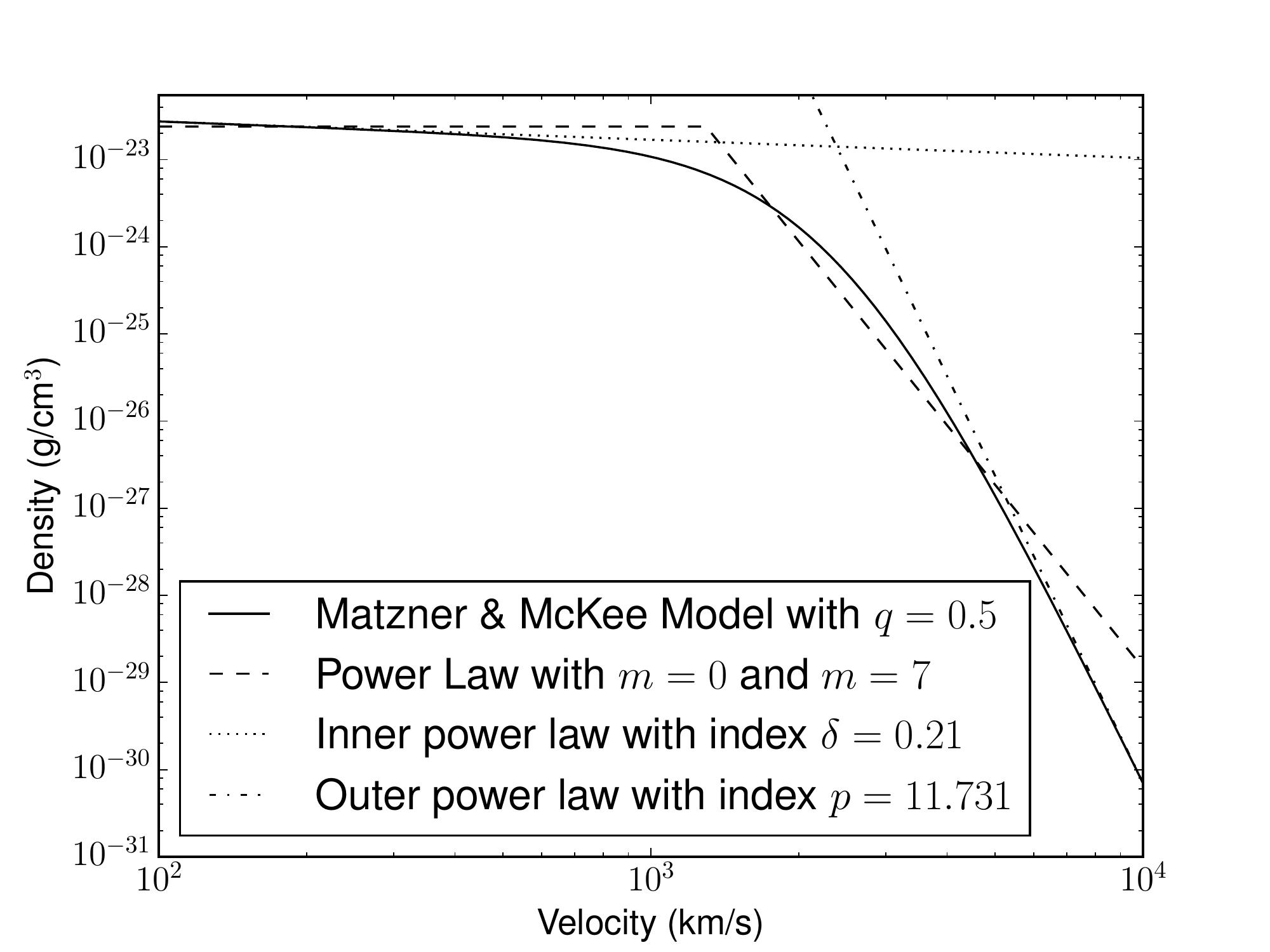}
\caption{Density distribution of the freely expanding supernova gas for an approximate power law structure and for a more detailed model of an exploded red supergiant star. The indices $\delta$ and $p$ represent the small velocity and large velocity asymptotes, respectively, for the detailed model.  
The approximate model with power law index $m$ is that given by equation (\ref{den}).
The assumed ejecta mass is 
$5\Msun$, the supernova energy is $10^{50}$ ergs, and the age is 940 yr.
}
\label{density}
\end{figure}

Our aim is to examine the evolution of the Crab Nebula, considering the dependence
on the supernova energy.
We begin with the supernova explosion, which results in an inner flat and outer steep density profile.
The ejecta are expected to reach free expansion within a few days and we approximate the profile by
\begin{equation}
\rho=\left\{
\begin{array}    
{ll}
At^{-3} & \mathrm{for}\, V\le V_t\\
At^{-3}\left(\frac{V}{V_t}\right)^{-7} & \mathrm{for}\, V>V_t,
\label{den}
\end{array}
\right.
\end{equation}
where $A$ is a constant, $V=r/t$ is the velocity of freely expanding gas, and $V_t$ is the transition velocity from a flat to a steep density profile.
A comparison to the more accurate profile used in Section 3.3 is shown in Figure \ref{density}.
The mass and energy of the supernova are given by $M_{\rm SN}=7\pi A V_t^3/3$
and $E_{\rm SN}=7\pi A V_t^5/5$, so that $V_t=(5E_{\rm SN}/3M_{\rm SN})^{1/2}=1.29\times 10^3 E_{50}^{1/2}M_5^{-1/2}\kms$,
where $E_{50}=E_{\rm SN}/(10^{50}{\rm~ergs}$) and $M_5=M_{\rm SN}/5\Msun$.
Also, $A=6.34\times 10^8 E_{50}^{-3/2}M_5^{5/2}$ g s$^3$ cm$^{-3}$.

On a  timescale of 100's of years, the pulsar bubble expands into the freely expanding
ejecta on the inside.
For the Crab pulsar, we take a period $P=0.0331$ s, $\dot P=4.2\times 10^{-13} $ s s$^{-1}$, and
a constant braking index $n=2.5$ \citep{lyne88}.
Using
\begin{equation}
P = P_0\left(1+\frac{t}{\tau}\right)^{\frac{1}{n-1}},
\label{pevol}
\end{equation}
where $P_0$ is the initial period and $\tau$ is a constant timescale, and the above parameters, we find $\tau =725$ yr for $t=940$ yr.
With  a neutron star moment of inertia of $10^{45}$ g cm$^2$, the current spindown power is $\dot E=4.6\times 10^{38}\ergs$
and it evolves as
\begin{equation}
\dot E = \dot E_0\left(1+\frac{t}{\tau}\right)^{-\frac{n+1}{n-1}},
\label{evol}
\end{equation}
where $\dot E_0$ is the initial power.
Assuming the observations were made at an age of 940 years, we have 
$\dot E_0=3.2 \times 10^{39}\ergs$.
The average power input up to time $t$ is
\begin{equation}
\dot E_a=\frac{(n-1)\dot E_0\tau}{2t}\left[1-\left(1+\frac{t}{\tau}\right)^{-2/(n-1)}\right].
\end{equation}
For the Crab pulsar parameters, $\dot E_a=0.39\dot E_0=1.2\times 10^{39} \ergs$
at the present time.

The expansion of the pulsar bubble is determined by the equations of conservation of
mass, momentum, and energy \citep{chevalier77}:
\begin{equation}
\frac{dM_{sw}}{dt}=4\pi R^2\rho (V_s-V),
\label{e1}
\end{equation}
\begin{equation}
M_{sw}\frac{d^2R}{dt^2}=4\pi R^2\left[P_i -\rho(V_s-V)^2\right],
\label{e2}
\end{equation}
\begin{equation}
\frac{d(4\pi R^4 P_i)}{dt}=\dot E R,
\label{e3}
\end{equation}
where $P_i$ is the interior bubble pressure and it has been assumed that the pulsar bubble expands adiabatically with an  adiabatic index $\gamma =4/3$.
The thin shell approximation is made in these equations.

The pulsar bubble initially expands in the constant density inner region of the ejecta.
If the pulsar power can be regarded as constant, the expansion is self-similar and
the bubble radius is \citep{chevalier77}
\begin{equation}
R=\left(\frac{875}{396}\frac{V_t^3 \dot E_a}{M_{SN}}\right)^{1/5}  t^{6/5}.
\label{rad}
\end{equation}
This evolution implies that the time to sweep out the ejecta to velocity $V_t$ is
\begin{equation}
t_t=\frac{132}{175}\frac{E_{SN}}{\dot E_a}.
\end{equation}
Taking $\dot E_a$ for the Crab pulsar, we have $t_t=1940 E_{50}$ yr,
so the shell is still primarily inside of the density inflection point.
Equation (\ref{rad}) can be used to find the velocity of the shell at the present age of the Crab
\begin{equation}
V_s=\frac{6}{5}\frac{R}{t}=1.28\times 10^3E_{50}^{3/10}M_5^{-1/2}\dot E_{a39}^{1/5}\kms.
\end{equation}
The velocity is less than that at the outer edge of the Crab Nebula, but is comparable to
that deduced from observations of bright filaments in the Crab.
An important point is that a supernova energy of $10^{50}$ ergs gives an outer velocity
in approximate agreement with the Crab Nebula dense filaments, while $10^{51}$ ergs gives too
high a velocity.

In this model, the ratio of swept up to total ejecta mass is
\begin{equation}
\frac{M_{sw}}{M_{SN}}=0.68\left(\frac{\dot E_a t}{E_{SN}}\right)^{3/5}=0.33E_{50}^{-0.6}\dot E_{a39}^{0.6}
\end{equation}
for the age of the Crab Nebula.
This value is at the low end of what is allowed by estimates of the observed swept up mass and the mass of the progenitor star.
Going to a supernova energy of $1\times 10^{51}$ ergs would decrease the predicted ratio
by a factor of 4 to well below the observed value.

In this simple model, $V_s=(6/5)(R/t_a)$, so that
\begin{equation}
\Delta T=t_a-\frac{R}{V_s}=\frac{t_a}{6}=157{\rm ~yr},
\label{delt}
\end{equation}
independent of other parameters.
The result is larger than that deduced from optical observations.
If there are the dependences $\dot E\propto t^{-s}$ and central density $A\propto r^{-\delta}$,
dimensional analysis shows that the $1/6$ factor in equation (\ref{delt}) becomes
$(1-s+\delta)/(6-s)$.
The density is expected to drop with radius, giving larger acceleration and an increasing
discrepancy between the theory and observation.
The power input $\dot E$ is expected to drop with time, decreasing the acceleration and
improving the discrepancy between the theory and observation.
The more detailed 
model presented in Section 3.3 has the capability to give better agreement with the observations.

The density in the flat part of the supernova density profile at an age 940 yr is
$\rho_f=2.4\times 10^{-23} M_5^{5/2} E_{50}^{-3/2}\gcm$.
If the free expansion velocity at the outer edge of the Crab is $1600\kms$, as discussed above, the density at that point is $0.54\times 10^{-23}\gcm$, which is somewhat below the preshock
density of  $2.1\times 10^{-23}\gcm$ inferred by \cite{sankrit97} in their radiative shock model.
An energy of $1.6\times 10^{50}$ ergs would improve the agreement, but with an energy of
$1\times 10^{51}$ ergs the highest ejecta density is $0.077\times 10^{-23}\gcm$.
The radiative shock model is more consistent with a low energy explosion.
The observed shell is only partial, at lower shell velocities \citep{sankrit97,hester08}, as would
be expected for the density profile proposed here.
We note that the outer edge is in the region of declining density here, whereas it is within this region in 
the discussion of the radial expansion because  the outer edge of the Crab Nebula is substantially outside of the
dense filaments.  This issue is further considered in Section 3.3.

\subsection{Interaction with Dense Mass Loss}

In the SN IIn-P model of \cite{smith13},
the supernova interacts with dense circumstellar matter, with mass $M_{CS}$, on a timescale of $\sim 10^2$ days.
The interaction may be responsible for the optical light from the supernova near maximum.
In the presence of the circumstellar material, the density profile in equation (\ref{den}) has a cut-off, at some velocity $V_c$. 
All the material with speed higher than $V_c$  interacts with circumstellar material and forms a 
dense shell with velocity $V_c$. 
We initially assume that the presupernova velocity of the initial circumstellar material is 
negligible, compared to the supernova expansion.   Under this assumption, the radial momentum of matter with velocity $>V_c$ is
$P_r(V>V_c)=[M(V>V_c)+M_{CS}]\times V_c$.
Assuming $V_c>V_t$, we find
$P_r(V>V_c)=M(V>V_c)\times 4V_c/3$,  leading to
$M(V>V_c)=3 M_{CS}$ after the circumstellar matter has been swept up.
The total mass of the shell, including circumstellar and ejecta components, is  $M_{Shell}=4 M_{CS}$.

For $V_c>V_t$ we find
\begin{equation}
V_c=\left(\frac{M_{SN}}{7M_{CS}}\right)^{1/4}V_t,
\end{equation}
so  the condition for $V_c>V_t$ is  $M_{SN}>7M_{CS}$.
When $V_c<V_t$,
an integral shows that the relation is
\begin{equation}
\frac{M_{CS}}{M_{SN}}=\left(\frac{V_c}{V_t}\right)^{-1}-1+\frac{1}{7}\left(\frac{V_c}{V_t}\right)^{3}.
\end{equation}
Values of $M_{SN}/M_{CS}$ for various values of $V_c/V_t$  are given in Table \ref{table}.

\begin{table}
\begin{center}

\caption{Supernova to circumstellar mass ratio}
\label{table}

\begin{tabular}{cc}
\hline
$V_c/V_t$ &  $M_{SN}/M_{CS}$  \\   
\hline
0.25 &  0.33   \\
0.5  &  0.98   \\
0.75 &  2.5   \\
1.0 &  7.0   \\
1.5 &  35   \\
2.0 &  112   \\
\hline
\end{tabular}
\end{center}
\end{table}

We have assumed here that the swept up shell of ejecta and circumstellar medium is thin because the internal
energy generated by the interaction is carried away by radiation.
This is a reasonable assumption in cases where the radiation is inferred to carry a
significant part of the supernova energy \citep[e.g.,][]{chugai04}.
The radiated energy is then
\begin{equation}
E_{rad}=\Delta E= E_{SN}(V>V_c)-\frac{1}{2} M_{Shell}V_c^2,
\end{equation}
where  $E_{SN}(V>V_c)$ is the energy in the freely expanding ejecta with velocities $>V_c$.
If $V_c>V_t$ we find
\begin{equation}
E_{rad}=\frac{5}{21}E_{SN}\left(\frac{V_c}{V_t}\right)^{-2}=\frac{25}{63}\frac{E_{SN}^2}{M_{SN}}V_c^{-2},
\end{equation}
so that $E_{rad}=4.0\times 10^{49} E_{50}^{2}M_5^{-1}V_{c3}^{-2}$ ergs, where
$V_{c3}=V_c/1000\kms$.
The value of $E_{rad}$ for SN 1054 is not well determined from the Chinese observations.
The light curve is roughly consistent with a normal Type II supernova, so that
$E_{rad}\approx (1-3)\times 10^{49}$  ergs.
\cite{mauerhan13} estimated a radiated energy of  $(2-3)\times 10^{49}$  ergs for 
the Type IIn-P SN 2011ht.
If $V_c=V_t$, we have $E_{rad}=2.4\times 10^{49} E_{50}$ ergs, which suggests that $V_c$
is close to $V_t$ or $1.29\times 10^3\kms$ for the standard parameters.
However,  the profile of freely expanding gas is cut off at $V_c$ and
we argued in Section 2 that there is freely expanding gas over the velocity range
$1600-2500\kms$.
The high velocity of free expansion suggests that $V_c$ is at least twice $V_t$, which drops $E_{rad}$
by a factor of 4.
There is thus a discrepancy between the Type IIn-P model for SN 1054 and the
evidence for freely expanding ejecta in the Crab Nebula.

In the model of \cite{tominaga13}, the progenitor is a super asymptotic giant (SAGB) star
with an estimated mass loss rate of $10^{-4}\ml$ and a wind velocity of $29\kms$.
In their discussion of SN 1054 as an electron capture event, \cite{moriya14}  took the
SAGB wind velocity to be $10\kms$.
The wind duration is $\sim 10^4$ yr, so it extends to $(3-9)\times 10^{17}$ cm,
or $(0.1-0.3)$ pc in radius.
Unlike the case of a Type IIn-P supernova, the forward shock wave is nonradiative from
an early age, although the reverse shock is radiative while the interaction region in is
the dense wind, for the standard parameters.
For the supernova density profile given in equation (\ref{den}), the interaction shell
moves through the wind on a timescale of 200 yr.
The velocity of freely expanding gas at that point is $1460\kms$.
In this case, the shocked gas adiabatically expands and cools after the shock
wave has traversed the dense wind, and the gas is ultimately expected to return to free expansion.
We do not examine the details of this process here, but note that energy conserving interaction
with a wind appears not to be ruled out for the Crab.

\subsection{Detailed Pulsar Nebula Model}

In the previous sections, we assumed a simple form for the freely expanding supernova density
profile as well as steady pulsar power input to enable analytic estimates of the model parameters.
Here, we use a supernova density profile based on detailed calculations \citep{matzner99} and
the standard evolution of the pulsar power (equation [\ref{evol}]).
The ordinary differential equations (\ref{e1}) -- (\ref{e3}) for the evolution of the pulsar bubble
are solved with a Runga-Kutta scheme. 

In previous discussions of pulsar nebula expansion in a supernova, the supernova has a normal
energy ($10^{51}$ ergs) and the expansion of the nebula is just in the inner flat part of the density
profile.
For a low energy supernova, the transition to the outer steeply dropping part of the
profile may become important.
The supernova density profile is shown in Figure \ref{density} for an energy of $1\times 10^{50}$ ergs
and an ejecta mass of $5\Msun$.
The model is intended for a red supergiant progenitor star, as might be expected
\citep{tominaga13}.
The model of \cite{matzner99} for red supergiants has a parameter $q$, which is 
related to the core-envelope mass ratio
and is in the range $0.3-0.8$; we choose $q=0.5$.
At small velocities, the profile approaches a flat power law with index $\delta=0.21$
and at high velocities a steep power law with index $p=11.7$ (Figure \ref{density}).

A problem with comparing the model to the actual Crab Nebula is that the model assumes
that the pulsar bubble sweeps up a thin shell, while the Crab filaments are spread in
radius, presumably as a result of the Rayleigh-Taylor instability.
We consider two cases to cover the possibilities.
In the first, the diameter of the Crab is taken to be 3.3 pc, consistent with the
outer boundary of the Crab.
Assuming that the filaments at the outer boundary have the typical $\Delta t=75$  yr,
we have a velocity of $1840\kms$.
For the other case, we assume that the velocity of the dense shell is characterized by
that of the dense filaments, about $1260\kms$ \citep{temim06}.
Now assuming that $\Delta t=75$ yr leads to a diameter of 2.23 pc.
We expect that the two cases will bracket the actual Crab Nebula.

\begin{figure}
\plotone{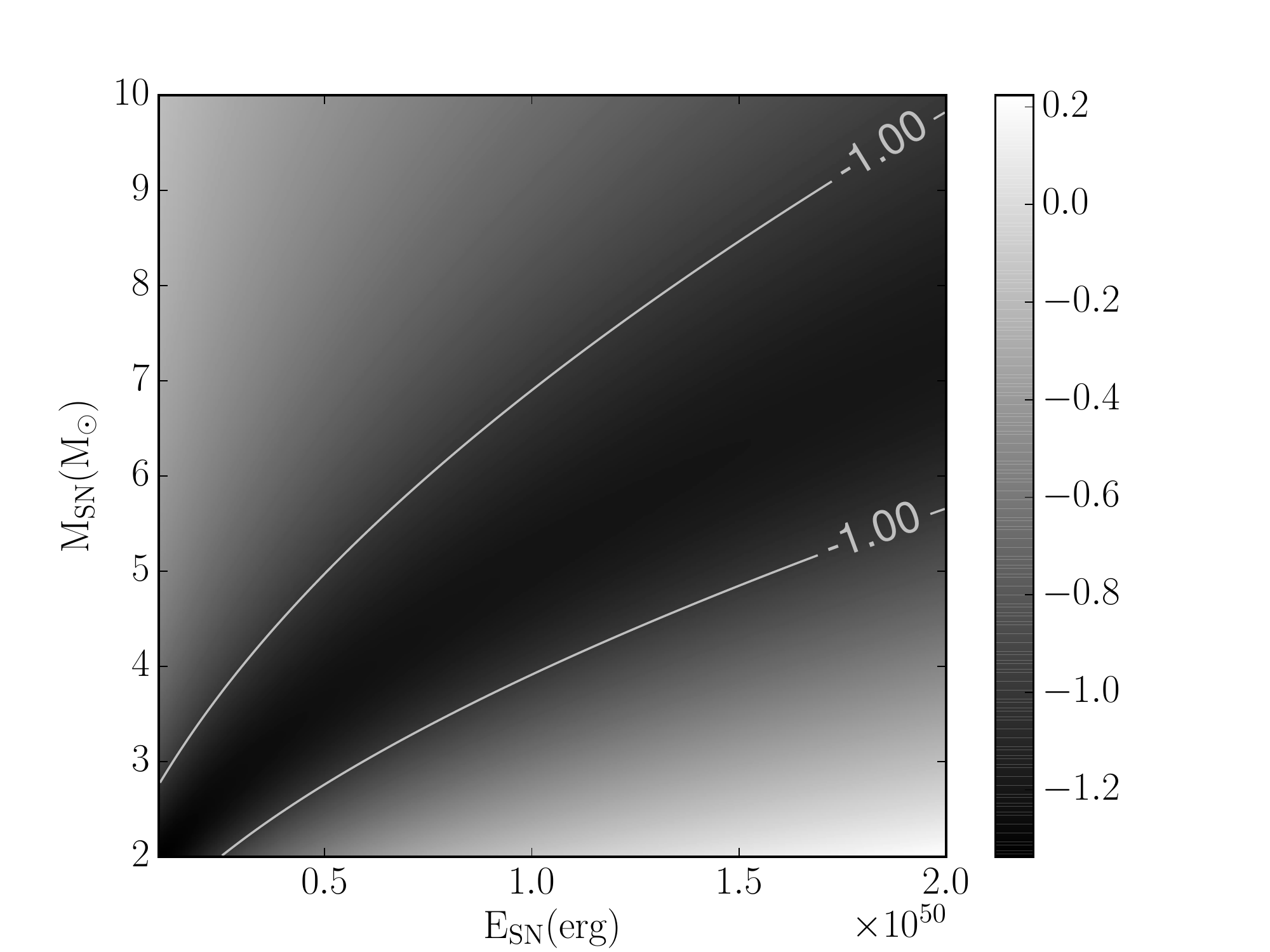}
\caption{Results of models as a function of explosion energy and total ejecta mass.
The shading gives the chi-squared value in fitting a shell radius of 1.65 pc and velocity
of $1840\kms$.}
\label{large}
\end{figure}

\begin{figure}
\plotone{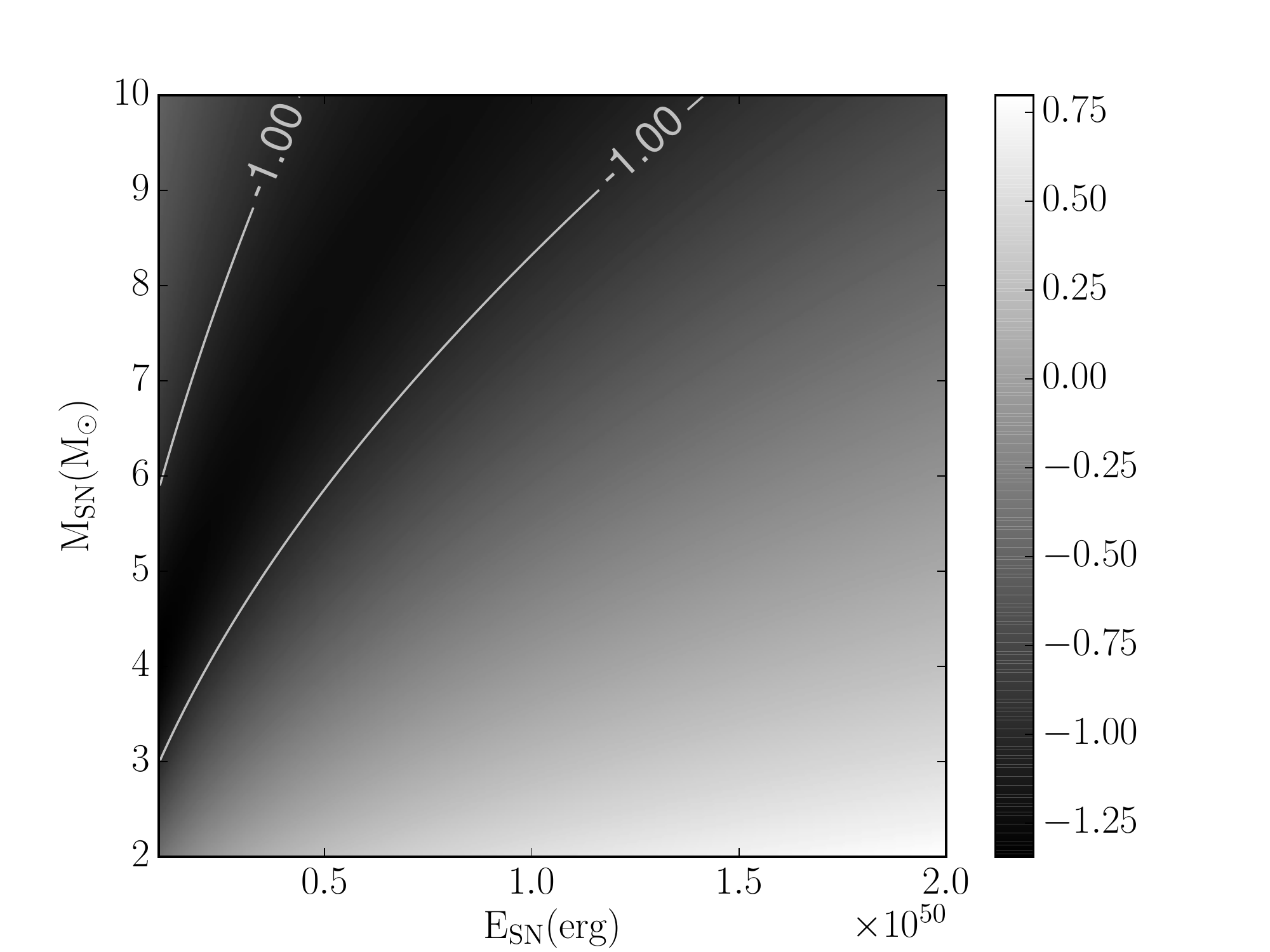}
\caption{Same as Figure 2, but fitting a shell radius of 1.1 pc and velocity
of $1260\kms$.}
\label{small}
\end{figure}

In order to estimate the goodness of fit of assumed models, we calculated
\begin{equation}
\chi^2 = [ (R_{fit} - R_0 ) / R_0 ] ^2 + [ (V_{fit} - V_0 ) / V_0 ]^2
\end{equation}
where $R_{fit}$ and $V_{fit}$ are the estimated observed values of the shell radius
and velocity, and $R_0$ and $V_0$ are the results of models for various values of
ejecta mass $M_{SN}$ and supernova energy $E_{SN}$.
In Figures \ref{large} and \ref{small}, the darkness of the plots indicates the value of $\log( \chi^2 )$.
Figure \ref{large} shows models with $R_{fit}=1.65$ pc and $V_{fit}=1840\kms$, where the model
parameters are the total ejecta mass, $M_{SN}$, and the supernova energy $E_{SN}$.
If $M_{SN}$ is $6.6\pm 1\Msun$, or less, as discussed in Section 2, the energy of the
supernova is $E_{SN}\approx 1.0\pm 0.5\times 10^{50}$ ergs.
Figure \ref{small} shows results for the assumption of the compact shell with
$R_{fit}=1.1$ pc and $V_{fit}=1260\kms$.
The acceptable models occur at low energy with $E_{SN}\la 0.5\times 10^{50}$ ergs.

\begin{figure}
\plotone{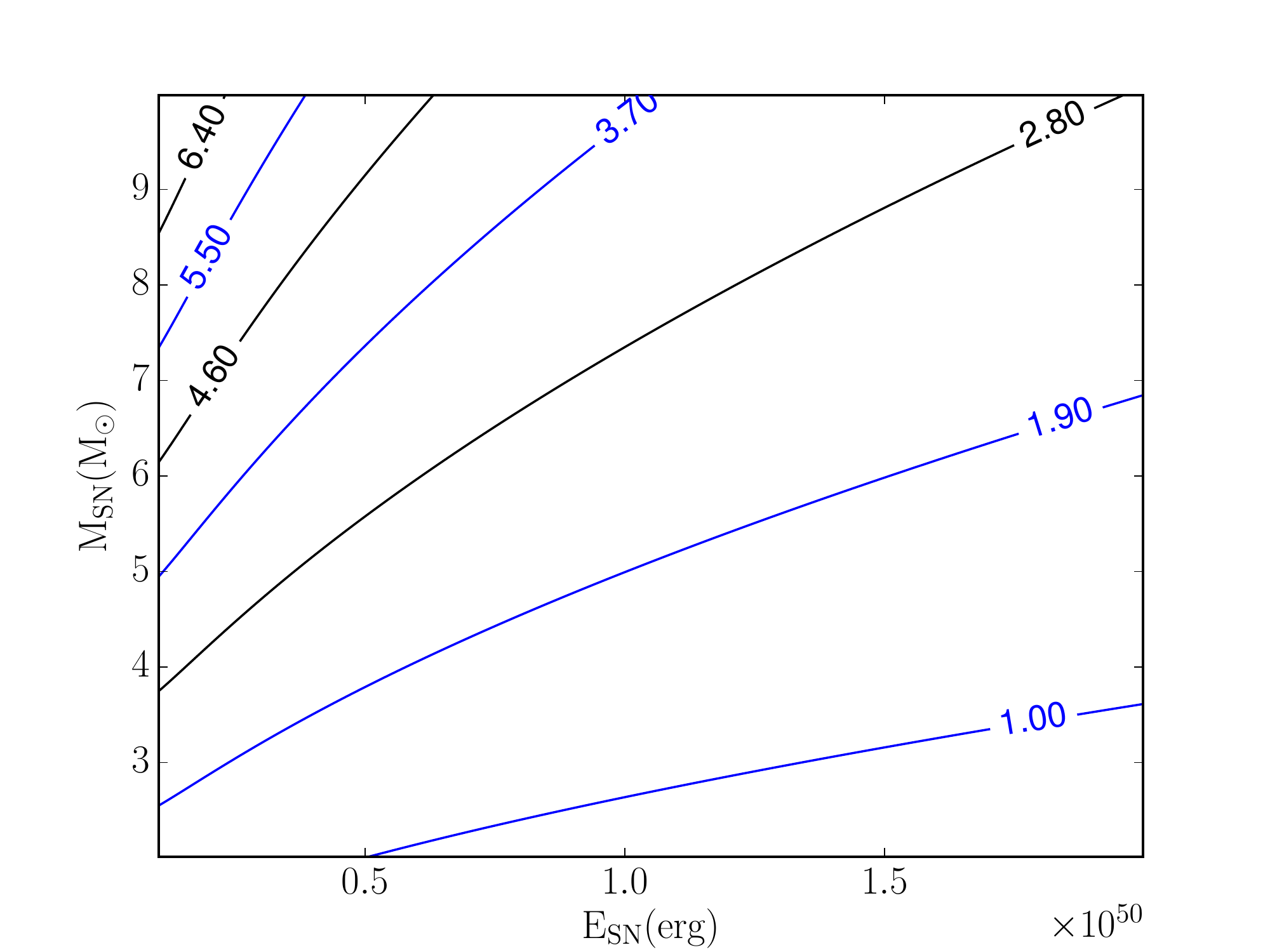}
\caption{The swept up shell mass resulting from a given explosion energy and total ejecta mass.
The labels on the solid lines give the swept up mass in $M_{\odot}$.
The black lines show $4.6 \pm 1.8 M_\odot$
where 1.8 $M_\odot$ is one standard deviation \citep{fesen97}; the blue lines show $+0.5$, $-0.5$, $-1.5$, $-2.0$ 
standard deviations, respectively. 
}
\label{mass}
\end{figure}

Another constraint comes from the swept up mass in the shell, which is estimated
as $4.6\pm 1.8\Msun$ from the observations \citep{fesen97}.
Figure \ref{mass} shows the swept up mass resulting from a given total ejecta mass, $M_{SN}$, and the supernova energy $E_{SN}$.
With the limit  on $M_{SN}$, a low value for the energy, $\la 10^{50}$ ergs, is
indicated, in agreement with the approximate model described in Section 2.1.
A recent estimate of the mass in the Crab filaments is $7.2\pm 0.5\Msun$ \citep{owen15}; using this
higher mass would yield an even stricter limit on the energy.

Also, the density profile of the freely expanding gas near the edge of the Crab Nebula (velocity
$\sim 1600\kms$) has
a density gradient between that of the outer steep power law and the inner flat region (Figure \ref{density}).
This is in accord with the findings of \cite{sollerman00} on the distribution of C IV
outside the main nebula.
A higher energy supernova would give a relatively flat density profile for the supernova
ejecta just outside the main body of the Crab.

\section{CONCLUSIONS}

Independent lines of evidence support a low energy ($\sim 10^{50}$ erg) for the supernova
giving rise to the Crab Nebula.
The evidence includes the radius and acceleration of the Crab shell,  the mass of swept up ejecta
in the Crab shell, the presence of a radiative shock wave around part of the Crab shell,
and the decreasing amount of C IV in the region surrounding the nebula.

The model proposed here combines the low energy supernova view of the Crab supernova
with expansion into freely expanding gas.
However, we do not consider the interaction of the freely expanding gas with its surroundings.
The nature of the presupernova mass loss and the interaction of the supernova with the
mass loss remains unclear.
Further studies of the region around the Crab Nebula are warranted.

\acknowledgments
We are grateful to Robert Fesen, Peter Lundqvist, and the referee for useful comments on the manuscript.
This research was supported in part by 
NASA grant NNX12AF90G.


\end{document}